# Directed High-Energy Infrared Laser Beams for Photovoltaic Generation of Electric Power at Remote Locations

Richard Soref[1, a)], Francesco De Leonardis[2], Gerard Daligou[3], and Oussama Moutanabbir[3]

[1]*Engineering Department, University of Massachusetts at Boston, Boston, MA 02125 USA*
[2]*Department of Electrical and Information Engineering, Politecnico di Bari, Bari, Italy*
[3]*Department of Engineering Physics, École Polyechnique de Montréal, Montréal, Québec, Canada*

[a)]**Author to whom correspondence should be addressed**: richard.soref@umb.edu

**ABSTRACT**
Transferring energy without transferring mass is a powerful paradigm to address the challenges faced when the access to, or the deployment of, the infrastructure for energy conversion is locally impossible or impractical. Laser beaming holds the promise of effectively implementing this paradigm. With this perspective, this work evaluates the optical-to-electrical power conversion that is created when a collimated laser beam illuminates a silicon photovoltaic solar cell that is located kilometers away from the laser. The laser is a CW high-energy Yb-doped fiber laser emitting at a center wavelength of 1075 nm with ~1 m$^2$ of effective beam area. For 20 kW illumination of a solar panel having 0.6 m$^2$ of area, optical simulations and thermal simulations indicate electrical output power of 3000 Watts at a panel temperature of 550 K. Our investigations show that thermo-radiative cells are rather inefficient. In contrast, an optimized approach to harvest laser energy is achieved by using a hybrid module consisting of a photovoltaic cell and a thermo-electric generator. Finally, practical considerations related to infrared power beaming are discussed and its potential applications are outlined.

## I. INTRODUCTION

Laser power converters for power-by-light and optical-wireless have been discussed in the literature [1,2], and this paper addresses the aspects of: (1) directed laser beams enabling electric-power generation at remote locations, (2) cases in which a very-high-power aimed beam travels through the ambient atmosphere to reach a targeted optical-to-electric (OE) converter that is located, for example, kilometers away from the laser, a remote location that is typically "off the grid". To leverage the atmospheric transparency windows, we are considering infrared lasers here, and not the visible ones.

The thesis of this paper is two-fold: (1) that powerful lasers have dual use for civilian and military purposes, (2) that the well-known silicon solar cell can also have dual use for harvesting laser beams as well as sunlight.

Using modeling, we have considered both thermo-radiative (TR) and photovoltaic (PV) receivers, and we find that the TR approach is much less efficient than PV in converting laser power into electric power. Our thermal simulations of the silicon solar PV cell show unavoidable heating of the cell, which traditionally is viewed as undesirable, but which we find beneficial in the laser case because the optimized electrical output occurs at a PV body temperature well above room temperature, such as 550K, as detailed in Section XII below. We have also investigated the combination of a PV cell with a thermo-electric generator (TEG) whose large area matches the area of the PV cell. Both PV and TEG are silicon-based and manufacturable. This PV + TEG hybrid provides electric power from both PV and TEG. The rear face of the TEG is ambient-cooled while the front TEG face thermally contacts the "heated" PV, thus providing a desired temperature-drop across TEG. We thus recommend PV+TEG as an optimized approach to directed-laser OE.

Our approach utilizes ultra-high-power (UHP) lasers whose optical output is 10 kW or more. In the UHP case, there is an interplay or tradeoff between the PV receiver spectral absorption, the temperature rise in the receiver and the receiver's OE conversion efficiency that decreases with increasing PV temperature, but a decrease that is not large-enough to prevent practical applications. Optical engineering and thermal engineering studies here allow us to select the optimum range of laser beam power. Optimizing the receiver load resistance is also important.

The sections of this paper cover UHP beaming lasers, potential and significant applications of beaming, silicon "solar" PV near-infrared application, PV conversion details, thermal and optical simulation results, benefits of using multi-solar-cell modules (panels), solar PV combined with group-IV TEG for high performance harvesting, analysis of the TR approach, and system costs-and-benefits.

## II. UHP LASER for BEAMING

We are considering infrared lasers rather than the ones that emit visible light, and we choose laser wavelengths corresponding to an atmospheric transparency window. Four relevant lasers within the near infrared, shortwave infrared, and longwave infrared satisfy this condition. 20 kW CW longwave infrared $CO_2$ lasers at the 10.6 μm wavelength are already commercially available for the present beaming applications

[3]. For the near infrared, there are commercial solid-state diode lasers whose output is anywhere within the 900 nm to 1080 nm wavelength range, and where 10 kW CW (or higher) output is available from several vendors [4-6]. Arrays of diodes are placed upon a bar, and then bars are stacked in several layers. The beams from stacked bars are then combined, giving a spot-focus with a condenser lens. To that spot, we would add a beam-collimating lens for our case. We also select two of these lasers, the 900 nm and the 1000 nm versions as having the highest transmission through the atmosphere. In addition, manufacturers have created the Er-doped fiber laser emitting at the 1567 nm wavelength, and this can be purchased in the 4 kW CW version [7]. Finally, we come to the most important or primary laser of this investigation, which is the Yb-doped fiber laser, that emits typically over a 1070 nm to 1080 nm band. Government agencies and contractors have developed UHP 50 kW-to-300 kW embodiments of these lasers for military applications [8]; but we are recommending here that such lasers have also potential applications in energy beaming. There is already a report of a 100 kW Yb fiber laser configured for cutting and welding applications [9]. And not least, there are 20 kW CW Yb-doped fiber lasers available commercially [10]. In this paper, our simulations cover the 0.1 to 50 kW laser power range, and we find laser power around 20 kW is optimal for energy harvesting.

Turning to the atmospheric transmission of laser beams, Figure 1 illustrates the high transmission of 1075 nm, 1567 nm and 10600 nm "directed-laser illuminators" by dots placed upon the atmospheric transmission spectrum.

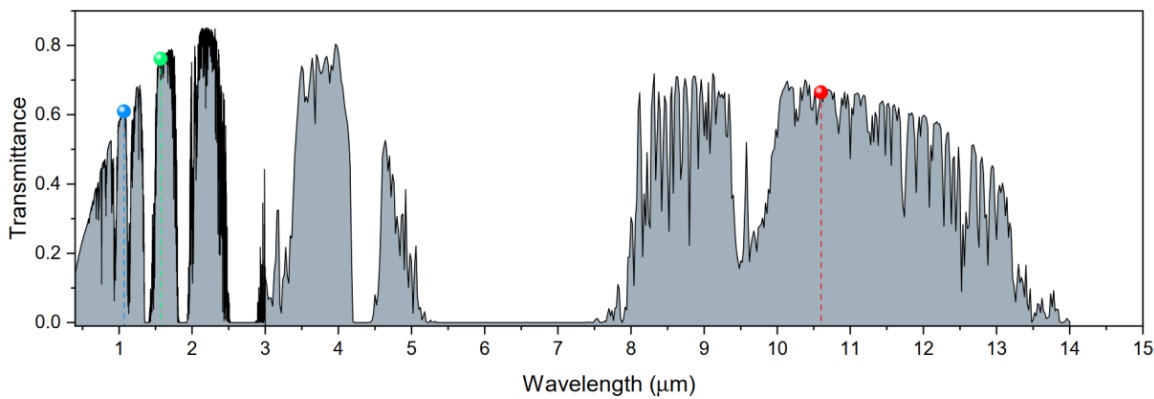

**FIG. 1.** Transmittance spectrum of the Earth's atmosphere at sea level [11]. The dots indicate the wavelengths of different lasers that can be considered for energy beaming: 1075 nm, 1567 nm and 10600 nm corresponding to high atmospheric transmittance.

We shall assume that the laser is in CW operation. Pulsed laser emission is feasible but is considered not optimum for electric power generation.

### III. PROPOSED APPLICATIONS and SCENARIOS

Using a distant target-mounted PV cell in conjunction with a powerful, pointed, laser beam can have various practical applications, especially in remote or off-grid areas. Here is a list of potential applications for the electric power generated by this proposed system.

*1*. Power generation for remote monitoring: This PV system would power remote equipment such as weather stations, wildlife cameras, or environmental sensors, in areas where access to the electrical grid is not feasible.
*2*. Wireless communication: The generated electricity can be used to power radio or satellite communication equipment, enabling communication in isolated areas, for example, during emergencies.
*3*. Renewable energy for off-grid homes and remote research stations: In unpopulated or remote areas with limited-or-no sunlight, PV cells powered by a distant laser can serve as a source of renewable energy for off-grid homes and cabins, providing electricity for lighting, appliances, and charging devices. Scientific research stations in remote locations with limited-or-no sunlight can benefit from the PV system to power scientific instruments, computers, and communication devices.
*4*. Pumping water for remote irrigation: The electricity generated can be used for pumping water in agricultural applications, providing a sustainable source of water for irrigation.
*5*. Emergency power for disaster relief: PV cells with a distant laser source can be deployed in disaster-stricken areas to provide emergency power for lighting, medical equipment, and communication devices.
*6*. Wildlife conservation efforts: The technology can support wildlife conservation efforts by powering cameras and tracking devices in remote regions.
*7*. Surveillance: Here the PV cells could be used for long-range surveillance and security applications, enabling continuous operation of cameras and sensors in strategic locations.
*8*. Remote power for space probes: In space exploration, PV cells could be used to power instruments and communication systems on unmanned spacecraft sent to distant planets or celestial bodies.
*9*. Remote powering of aircraft: Assuming that the beam-pointer tracks the aircraft in real time, the aircraft-mounted PV can power that airplane. This application overlaps the DARPA Persistent Optical Wireless Energy Relay program announced in 2023, a program in which the goal is to mount the UHP laser on a flying aircraft and to beam that power over distances up to 200 km to other flying aircraft.
*10*. Environmental monitoring stations: PV-powered stations can continuously monitor environmental conditions in remote areas, helping to collect valuable data for research and conservation efforts.

*11.* Mining and resource extraction: In remote mining or resource extraction operations, PV cells can provide electricity for essential equipment and/or communication systems.

When implementing these systems, it is important to consider factors such as laser safety, PV opto-electric efficiency, local regulations, environmental considerations, and the costs of deploying and maintaining the system. Atmospheric issues to contend with include thermal blooming of the beam, atmospheric turbulence, smog, fog, smoke and rain. As indicated, the laser can be on earth, shipborne, airborne, or space-borne.

## IV. PHOTOVOLTAIC OE CONVERTERS OF LASER ENERGY

We have investigated the PV approach, the hybrid PV-and-TEG approach and the thermo-radiative (TR) diode approach to OE conversion, detailed as follows. We are proposing direct illumination of the PV diode's input face by the laser as being the most effective situation. One can also envision an indirect PV approach in which a sheet or thick layer of absorber-emitter (a-e) material is placed in front of the PV diode's front side in order to absorb laser light in the a-e, thereby heating it, and thereby creating a blackbody emitter at some high temperature that then radiates to the "nearby" PV diode. However, even if we assume that the PV absorbs a large portion of the blackbody radiation spectrum, this a-e scenario is not optimum for OE conversion because the blackbody power density in W/m$^2$ (averaged over the absorbed spectral region) is a smaller power density than that supplied by the UHP laser, even for high blackbody temperatures such as 1000 K or 1200 K. In summary, the direct PV illumination has obviously higher efficiency.

## V. SILICON SOLAR CELL for NIR BEAMED OE

The OE conversions of the 1570 mm and 10600 mm UHP laser beams will be quantified in a subsequent study, while in this work we shall examine the near-infrared cases. Considering PV semiconductors generally, PV theory indicates that the optimum bandgap wavelength $\lambda_g$ of the PV diode is slightly longer than the laser wavelength. For the 900-to-1000 nm and 1075 nm UHP lasers, it is fortuitous that the extremely familiar silicon PV solar cell ($\lambda_g$ = 1107 nm at 300K) satisfies this relation. Hence efficient OE conversion is expected, as we quantify.

The direct-illumination approach is shown in Fig. 2, and here it is important to remove heat from the PV which is done to some extent by the metal heat sink at the rear, which works in the ambient air. There is also air convention cooling at the front and the cell emits gray body radiation with 0.8 emissivity. Fig. 3 presents the cross-section view of the most popular solar cell [12,13]. It is seen that the absorption of the laser beam takes place across the 180 μm (or 500 μm) thickness of the cell.

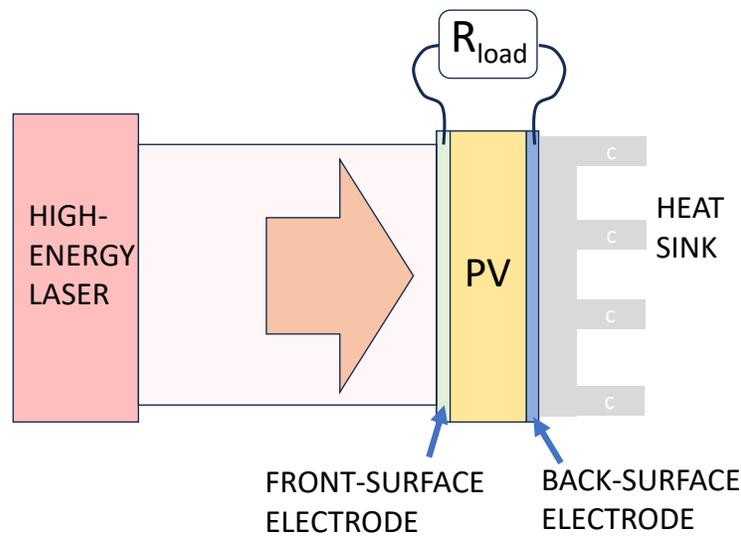

**FIG 2.** Schematic view of laser power beaming by means of a PV cell.

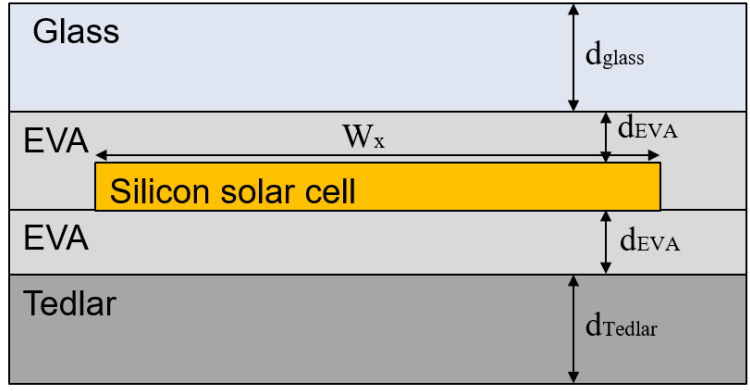

**Fig. 3.** Cross section view of high-performance cost-effective Silicon Solar Cell. Relevant parameters and materials are indicated. A standard commercial cell has an area of 15.6 cm x 15.6 cm.

## V. THERMAL ANALYSIS OF THE Si PV APPROACH

In this section, the thermal physics model of the laser heating of PV structures is presented. Here, the purpose is to describe the main effects in order to ascertain the PV figure of merit as a function of the PV body temperature, induced by laser illuminations.

The PV is irradiated by means of a laser spot having a size opportunely chosen to induce the heating effect over most of the PV cell area. In this context, the three-dimensional energy equation of the PV structure is expressed by Eq. (1), where $C_p$ is the specific heat capacity, $k$ is the thermal conductivity and $\rho$ is the density; $T$, $Q_{laser}$, and $Q_{PV}$ represent the temperature, the heat source induced by the laser-energy absorption and the electrical power density generated by the photovoltaic volume, respectively.

$$\rho C_p \frac{\partial T}{\partial t} + \rho C_p \mathbf{u} \cdot \nabla T - \nabla(k\nabla T) = Q_{laser} - Q_{PV} \quad (1)$$

Eq. 1 clearly evidences that the temperature $T$ of the system is related to the difference between the heat source $Q_{laser}$ generated by the absorption of the laser and the internal heat sink induced by $Q_{PV}$. According to the Beer-Lambert law, the heat source in each layer ($Q_{laser}^{(i)}$) is given by:

$$Q_{laser}^{(i)} = Flux^{(i-1)} \cdot \alpha_i \cdot e^{(-\alpha_i \cdot z)} \quad (2)$$

Where the subscript $i = 1,2,3,4,5$ stands for glass ($i = 1$), EVA ($i = 2$, and 4), Si ($i = 3$), Tedlar ($i = 5$) and $\alpha_i$, is the absorption coefficient in the i$^{th}$ layer. The term $z$ represents the coordinate in the vertical direction of the PV cell. Since the absorption coefficient of the silicon is dominant, during the entire laser illumination, the instantaneous energy of the laser beam is mainly absorbed by the silicon layer (see Fig. 3) and is converted into a thermal source.

Moreover, the silicon loss coefficient $\alpha(\lambda, T)$, dependent on the temperature and wavelength, is calculated as [14]:

$$\alpha(\lambda, T) = \sum_{\substack{i=1,2 \\ j=1,2}} C_i A_j \left\{ \frac{[\hbar\omega - E_{gj}(T) + E_{pi}]^2}{\left[e^{\frac{E_{pi}}{k_B T}} - 1\right]} + \frac{[\hbar\omega - E_{gj}(T) - E_{pi}]^2}{\left[1 - e^{\frac{-E_{pi}}{k_B T}}\right]} \right\} +$$

$$A_d \left(\hbar\omega - E_{gd}(T)\right)^{1/2} \quad (3)$$

where the indirect ($E_{gj}(T)$) and direct ($E_{gd}(T)$) bandgap energies are expressed as a function of temperature by means of the Varshni equations [14]. Moreover, the values of the phonon energies ($E_{pi}$) and of the fitting constants $C_i$, $A_j$, and $A_d$ are listed in Table I of Ref. [14].

By assuming that the laser beam has a Gaussian spatial distribution, the laser flux in the top is given by:

$$Flux = (1 - R) \frac{2 \cdot P_{laser}}{\pi \cdot r_{sp}^2} e^{\left(-2 \frac{(x - x_{focus})^2 + (y - y_{focus})^2}{r_{sp}^2}\right)} \quad (4)$$

Where $R$ is the reflectivity at the top glass surface, $P_{laser}$ is the input laser power and $r_{sp}$ represents the laser spot radius. The terms $x$ and $y$ are the coordinates in the xy plane of the PV cell and $x_{focus}$ and $y_{focus}$ represent the spot's center point.

In order to realize a self-consistent model, the internal heat sink $Q_{PV}$ is calculated as $Q_{PV} = \eta \times Q_{laser}^{(Si)}$, where $\eta$ represents the PV conversion efficiency for monochromatic illumination. In this context, Green *et al*. [15] have experimentally demonstrated conversion efficiency in silicon cells above 45%, under illumination equivalent to monochromatic light intensities of about 1 W/cm$^2$. In particular, efficiencies close to 40% were recorded for light at the 1064 nm wavelength as emitted by neodymium-doped yttrium-aluminum garnet (Nd:YAG) lasers. In this sense, we think that Green's theoretical derivation (summarized in the following) offers good agreement with the experimental data and can be well integrated with the 3D thermal simulations. Thus, the conversion efficiency for monochromatic illumination is evaluated as:

$$\eta = \frac{J(V)V}{Flux} \quad (5)$$

where the density current depending on the voltage $V$ is calculated as:

$$J(V) = q \int_{E_g}^{\infty} a(E) n_{in}(E) dE - q \frac{2\pi \Delta\Omega}{h^3 c_0^2 ERE} \int_{E_g}^{\infty} a(E) \frac{E^2}{\left[e^{\left(\frac{E-qV}{kT}\right)}-1\right]} dE \quad (6)$$

Where $\Delta\Omega$ indicates the solid angle. The first term in Eq. 6 is the absorbed photon flux from the laser beam. The photon-energy dependent coefficient $a(E)$ is the PV absorbance and $n_{in}(E)$ represents the input photon flux density, assumed to have a Gaussian distribution around the central photon energy emission with FWHM $\delta E = 2\pi c_0 \delta\lambda/\lambda^2$, where $\delta\lambda$ is the emission bandwidth. Finally, the term $ERE$ (external radiative efficiency) represents the fraction of all photon losses from the device (i.e. nonradiative recombination).

## VI. RESULTS of THERMAL and OPTICAL SIMULATIONS

We investigated the thermal aspects of the laser illumination of the silicon PV sketched in Fig. 3, and we modeled the laser heating using the Finite Element Method (COMSOL Multiphysics) models. We performed 3D simulations where the heat-transfer-in-solids model is coupled together with the model of the conversion efficiency (see Eqs. 5 and 6) in an integrated approach to include in the thermal simulations (see Eq. 1) the internal heat-sink effect induced by the electrical power generation. Some main experimental parameters such as the laser power, laser spot size, and the thickness of the silicon cell are discussed in detail to investigate their influence on both the temperature distribution and the PV figure of merit.

According to Fig. 3, we targeted the silicon PV having $W_x \times W_y$ =15.6 cm × 15.6 cm, $d_{glass} = 3$ mm, $d_{EVA}$ =0.4 mm and $d_{Tedlar} = 0.5$ mm. Moreover, in the following analysis, we assumed the silicon layer thickness, $d_{Si}$, of 180 μm and of 500 μm, and the wavelength emission of the UHP laser $\lambda$=1075 nm, with the emission bandwidth $\delta\lambda$=10 nm.

As a first step, we performed parametric simulations based on Eqs. 5 and 6 in order to evaluate the influence of the design parameters on the conversion efficiency. In this context, our investigations are plotted in Figs. 4 and 5. In particular, Fig. 4 shows the conversion efficiency as a function of the voltage, for different values of the laser power, where the operative temperature is controlled at 300 K. As is well-known, the output electrical power density, $J(V)V$, must be optimized over $V$. As a result, the maximum conversion efficiency (at the Maximum Power Point, MPP) is evaluated at the optimum $V$ and is then adopted in the following analysis as the figure of merit. The curves of Fig. 4 clearly indicate that the maximum conversion efficiency increases by increasing $d_{Si}$ from 180 μm to 500 μm, with a slope $\partial\eta/\partial d_{Si}$= 0.0004418 μm$^{-1}$. Since the operative temperature is forced to 300 K, this trend is essentially depending upon the different values of the PV absorbance. Indeed, because the calculated absorption depth is 940 μm (at 300 K), a larger number of photons are absorbed when $d_{Si}$ is ~500 μm, resulting in a larger generated density current $J(V)$.

Using the optimum $V$ value, we then calculate the maximum conversion efficiency as a function of laser spot radius ($r_{sp}$), presented in Fig. 5. In that Figure, the laser power is assumed as 10000 W or 50000 W and the operative temperature is controlled to 300 K. Related to the absorbance, a strong linkage of $\eta_{max}$ to $d_{Si}$ is found in Fig. 5, as it was in Fig. 4. The four curves in Fig. 5 exhibit a modest decrease of $\eta_{max}$ with increasing spot radius, and the cause of this is the decrease in the Gaussian-distributed photon flux density $n_{in}(E)$ in Eq. (6).

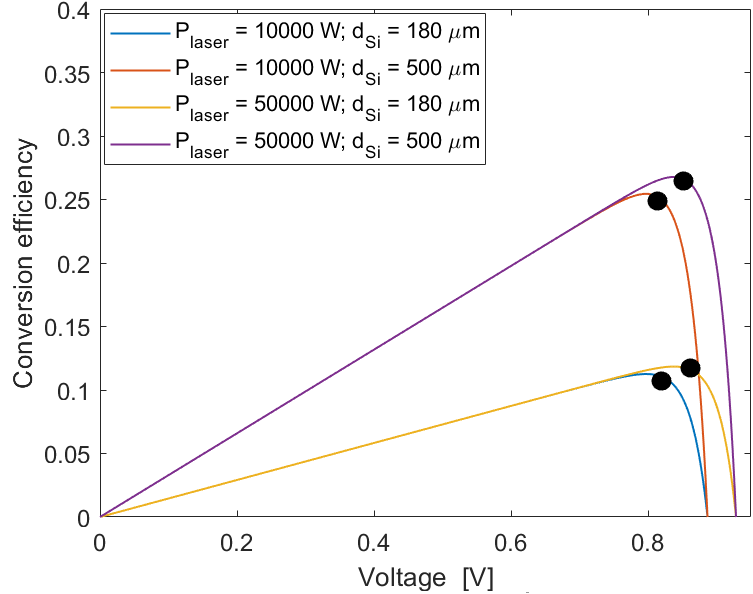

**Fig. 4.** Conversion efficiency as a function of the voltage, for laser power values of 10,000 W and 50,000 W and $d_{Si}$ =180 μm and 500 μm, respectively. In the simulations, the laser emission wavelength and the laser spot radius ($r_{sp}$) are 1075 nm and 80 cm, respectively. The operative temperature is forced at 300 K.

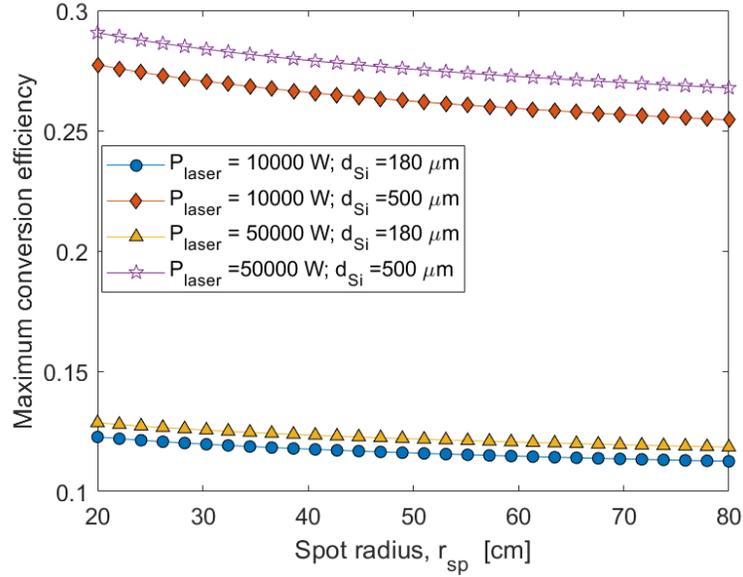

**Fig. 5.** Maximum conversion efficiency as a function of the laser spot radius, for laser power values of 10000 W and 50000 W and $d_{Si}$ =180 µm and 500 µm, respectively. In the simulations, the laser wavelength emission and the emission bandwidth are 1075 nm and 10 nm, respectively. The operative temperature is forced at 300 K.

TABLE I. PHYSICAL PARAMETERS OF THERMAL SIMULATION

| Parameters | Materials | | | |
| --- | --- | --- | --- | --- |
| | Silicon | Glass | EVA | Tedlar |
| Density [kg/m$^3$] | 2330 | 2450 | 950 | 1200 |
| Thermal conductivity [W/mK] | 130 | 2 | 0.311 | 0.15 |
| Heat capacity at constant pression [J/kgK] | 677 | 500 | 2090 | 1250 |

From the plot, we record that the maximum conversion efficiency decreases with increasing spot radius with a slope of $\partial\eta/\partial r_{sp}$ = - 0.0001598 cm$^{-1}$ and -0.0003613 cm$^{-1}$ for $d_{Si}$=180 µm and 500 µm, respectively. Moreover, the curves show that $\partial\eta/\partial r_{sp}$ values are weakly dependent upon the level of the laser power.

At this step, we present the heating effect upon the maximum conversion efficiency, by performing 3D simulations integrating Eqs. 3-6 in the FEM tool. The material properties used in the thermal simulations are summarized in Table I.

The conditions considered in the thermal simulations are as follows: (a) the thermophysical parameters of all photovoltaic materials are presumed to be isotropic and independent of temperature, (b) The PV side boundary is considered adiabatic, (c) The initial temperature of the PV structure is equal to the ambient temperature ($T_0$=293 K), (d) the heat flux due to heat losses by radiative and convective heat transfer between the PV and environment are applied to the top and bottom surfaces.

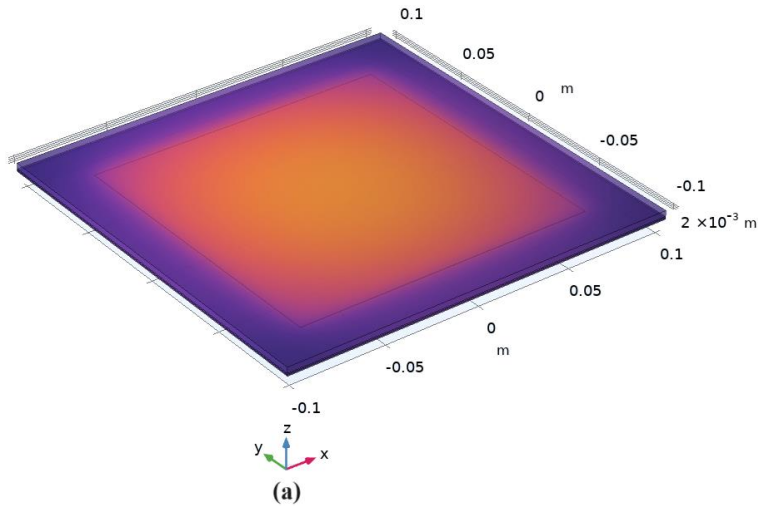

(a)

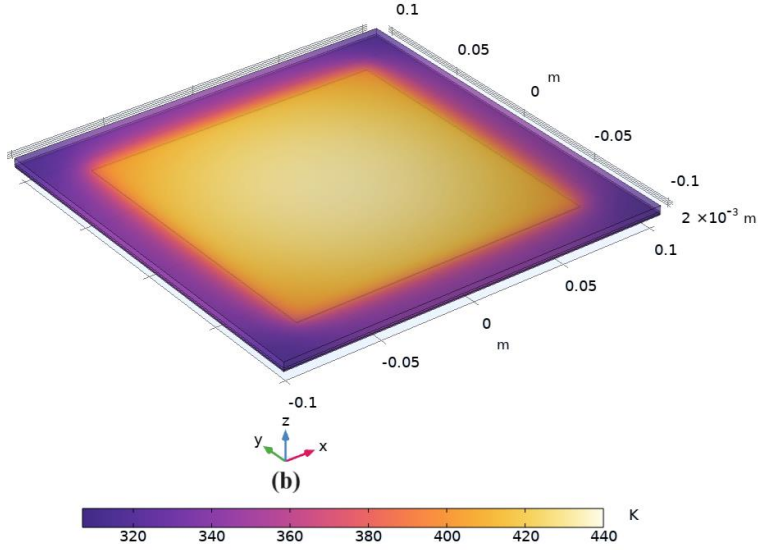

**FIG. 6.** Temperature distribution in a 156 mm x 156 mm silicon PV. (a) Laser power 10000 W and $d_{Si}$ =180 µm; (b) Laser power 10000 W and $d_{Si}$ =500 µm. In the simulations, the laser wavelength emission, the emission bandwidth and the laser spot radius ($r_{sp}$) are 1075 nm, 10 nm and 80 cm, respectively.

Regarding convection, we have assumed here a passive heat sink at the PV bottom surface, a sink consisting of a fin array that is exposed to the ambient air in order to dissipate heat. Although this convection cooling is helpful, it does not prevent PV temperature rise. If we had imposed, instead of convection, the constraint that the body of the PV converter must always be held at 293K during the incoming UHP beaming, that would have required considerable *active* heat sinking, which means that electrically powered devices would be deployed: blowers to force air onto the sink, or pumps to send fluids through the sink. Such electric powering would in our view be counter-productive since it would subtract significant power from the "net" electric power generation. For example, it would be a formidable electrical task to remove the heat induced by 20 kW of PV-absorbed power.

In this context, the spatial temperature distribution for the laser heating process is shown in Figs. 6 (a) and (b), for $d_{Si}$=180 µm and 500 µm, respectively. In the simulations a laser power of 10000 W has been assumed. The plots indicate clearly that the silicon PV with $d_{Si}$=500 µm reaches higher temperature values. This trend can be explained in terms of the absorption depth. When the temperature increases, the silicon absorption increases and then the absorption depth reduces. We record the absorption depth values of 940 µm decreasing to 253 µm when changing the temperature from 300 K to 400 K. Under these conditions, the number of absorbed photons is maximized for $d_{Si}$=500 µm with respect to the case $d_{Si}$=180 µm, inducing a larger heat source term in Eq. 2.

Using Eqs. 2-6, and taking into account the increased PV absorption induced by bandgap shrinkage in silicon when the body temperature increases above 293K, we have plotted in Figure 7 the resulting conversion efficiency-versus-temperature for incoming laser power at the 905 nm, 950 nm, 1000 nm, 1060 nm and 1075 nm wavelengths.

Our parametric investigation of the 1075-nm laser heating process is given in Fig. 8 which shows the maximum conversion efficiency (left axis) and the PV body temperature (right axis) as a function of the laser power, for $d_{Si}$=180 µm and 500 µm, respectively. In both cases, the maximum conversion efficiency reaches a peak value ($\eta_{peak}$), corresponding to a particular value of the laser power ($P_{peak}$) to which corresponds, in turn, a well-determined heating temperature ($T_{peak}$).

It is worth outlining that, in our integrated approach, the temperature dependence of the conversion efficiency is taken into account by means of the following main contributions: i) the density current relationship of Eq. 5, ii) the absorbance by means of the silicon absorption coefficient (see Eq.3), iii) the Varshni equation applied to the silicon energy band gap [14].

At this point, we will assert that the electrical output power density is the most practical or meaningful figure of merit for this laser beaming, and for that reason we have simulated the density increase with laser power, as given in Fig. 9 (left axis) together with results for the associated PV body temperature (right axis). Both density and temperature increase strongly with beam power. Now we shall, somewhat arbitrarily, set a limit of 550K as being the maximum useable operating temperature of the Si PV. With that provision, we have obtained the OE results summarized in Table II for incident laser powers of 10000, 15000, 19013, and 19388 W, respectively.

Regarding the OE figure of merit for Table II, the traditional approach is to use the power conversion efficiency $\eta_{max}$ at the maximum power point of the conversion device. This merit figure is given by $\eta_{max} = D_c/D_b$ where we define the beam density $D_b = P_b/A_b$, where $P_b$ is the input power of the laser beam, and $A_b$ is the effective area of the collimated incident laser beam, and where $D_c = P_c/A_c$, in which $P_c$ is the output electric power of the cell, and $A_c$ is the area of that cell. We assume a collimated cylindrical laser beam with Gaussian intensity distribution, for which $r_{sp}$ is the radius where the beam intensity has fallen down to 1/e² of the central beam power. Then we define the effective area of the beam $A_b = \pi r_e^2$, where $r_e$ is the effective beam-spot radius given by $r_e = r_{sp}/\sqrt{2}$, yielding $A_b = \pi r_{sp}^2/2$. In Table II, the effective area of the Gaussian laser column is one square meter, based upon the beam-column effective radius of 56.6 cm.

In practice, the most important figure of merit is the absolute electric power emerging from the cell, which is $P_c = \eta_{max} P_b A_c / A_b$. Using the $P_c$ criterion, the electric output power results in Table II are 146 W ($d_{Si}$=180 µm) and 157 W ($d_{Si}$=500 µm) for input laser power of 19013 W and 19388 W, respectively. These electrical results refer to a solar cell whose area is only 2.4 % of the effective area of the incident beam. This immediately indicates that we require larger-area OE converters to achieve larger electrical outputs.

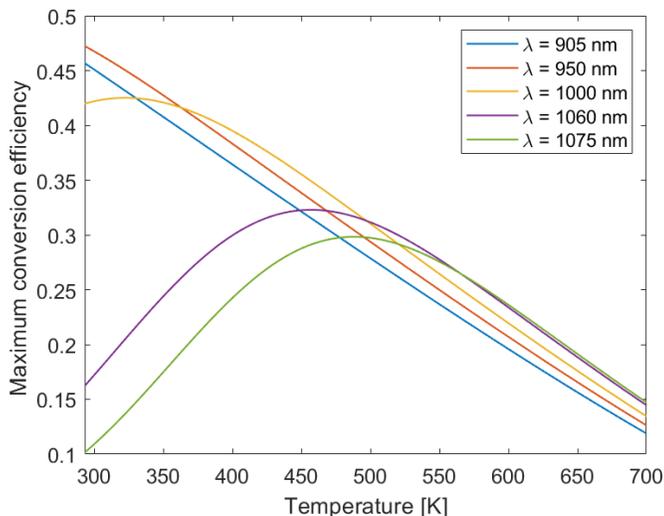

**Fig. 7.** Maximum conversion efficiency as a function of the Si PV body temperature for five laser-emission wavelengths. In these simulations, $d_{Si} = 180$ µm.

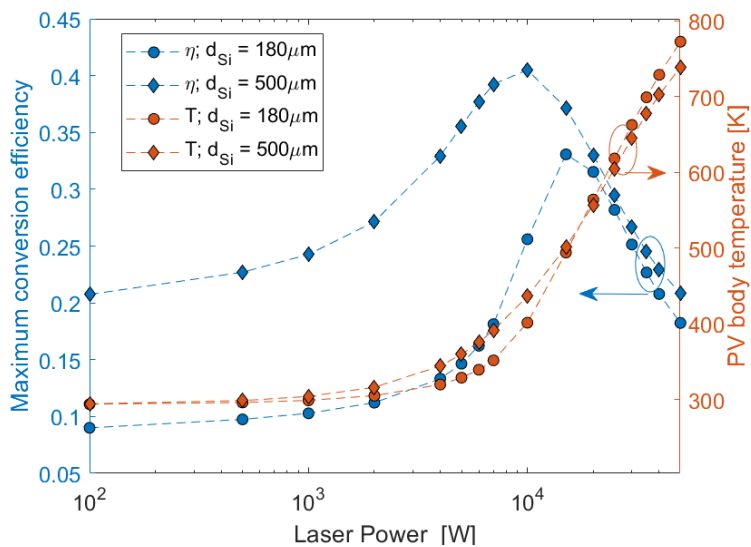

**FIG 8.** Maximum conversion efficiency and PV body temperature as a function of the laser power, ranging from 100 W to 50000 W and for $d_{Si}$ =180 µm and 500 µm. In the simulations, the laser wavelength emission, the emission bandwidth and the laser spot radius ($r_{sp}$) are 1075 nm, 10 nm and 80 cm, respectively.

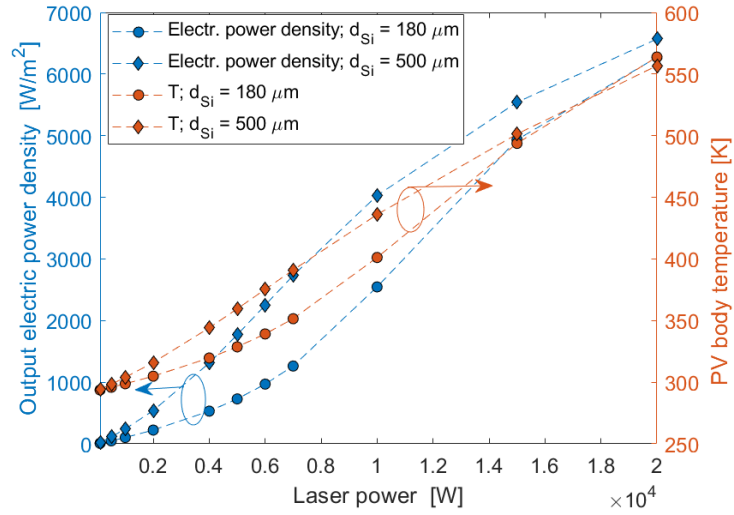

**FIG 9.** Output electrical power density and its associated PV body temperature as a function of the laser power, ranging from zero W to 20000 W and for $d_{Si}$ =180 μm and 500 μm. In the simulations, the laser wavelength emission, the emission bandwidth and the laser spot radius ($r_{sp}$) are 1075 nm, 10 nm and 80 cm, respectively.

TABLE II. SUMMARY OF RESULTS

| Laser power input [W] | Metrics | | | |
|---|---|---|---|---|
| | PV body temperature [K] | Efficiency | Output electric power density [W/m²] | Output electric power [W] For A=0.0243 m² |
| **19013** ($d_{Si}$=180 μm) | 550 | 0.32 | 6008 | 146 |
| **19388** ($d_{Si}$=500 μm) | 550 | 0.33 | 6443 | 157 |
| **15000** ($d_{Si}$=180 μm) | 501 | 0.33 | 4935 | 120 |
| **10000** ($d_{Si}$=500 μm) | 436 | 0.41 | 4029 | 98 |

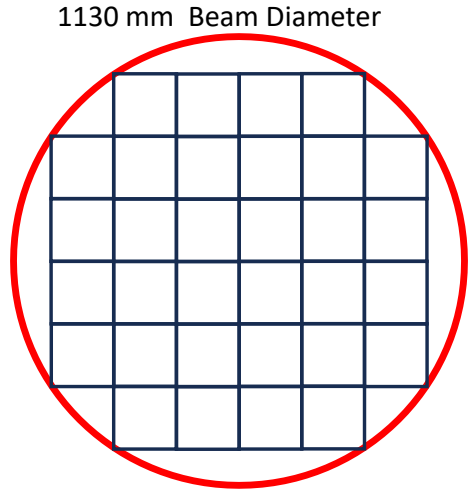

**Fig. 10.** Schematic drawing of 32-cell Si PV panel having area of 0.78 m². Array overlap with the laser beam is shown.

## VII. SOLAR PANEL ESTIMATES

If we now consider a 5 x 5 interconnected array of the above-cited standard cells in order to construct a custom-made solar panel, then the PV area increases to approximately 0.61 m² which is a factor-of-25 enlargement. Taking that panel as the new OE converter, we then propose to use the same 1.0 m² effective-area cylindrical laser beam ($A_b$) to illuminate and flood the panel. The electrical power output $P_c$ scales as the PV area $A_c$ according to the above relation. $P_c = \eta_{max} P_b A_c / A_b$. One estimate is that $P_c$ will increase by a factor of 25, but a more conservative approach is to say that $P_c$ will be higher by a factor of 15 or of 20 because the beam intensity is not constant across its area $A_b$ (nonuniform illumination). In that case, using the factor-of-15 and factor-of-20 predictions, taking the Table-II result of 157 W at 500 μm Si PV thickness (550K PV body), we find for the 15 and 20 predictions $P_c$ = 2355 W and 3140 W, respectively. Therefore, there appears to be a realistic pathway to 3000 W output for 20000 W laser input. We should mention that the 32-module array proposed in Figure 10 appears to give a 28% improvement over the 5 x 5 array.

We considered factors that could limit the performance of a multi-module panel; arrays such as 5 x 5 or 32 or 6 x 6. Specifically in Section VI above, and in Table II, for an individual module we took into account the external radiative efficiency (ERE), a parameter that accounts for carrier recombination losses. Resistive losses were not included. However, resistive effects in individual solar cells do reduce the fill factor FF and then reduce the efficiency of the cell by dissipating power in the resistances. In particular, the most common parasitic resistances are the series resistance and the shunt resistance. In our PVs modules, typical values for area-normalized series resistance are around 0.5 $\Omega cm^2$. By contrast, the values for the shunt resistance are in the M$\Omega cm^2$ range for laboratory-type solar cells, and are 1000 $\Omega cm^2$ for commercial solar cells. In this context, due to the large area used for one cell (0.0243 $m^2$), the dominant resistive effect is determined by the series resistance $R_{SER}$.

To accurately represent our case, we shall make the assumptions $d_{Si}$=500 µm, laser power of 10000 W and PV body temperature of 436 K. Then we estimate that the series resistance induces an electric power loss of around 52% for the single PV cell. However, considering the load resistance connected to the panel, the resistive effect has a negligible effect on a series-interconnected multi-module panel array. Thus, we propose to connect the entire array in electrical series. The result of that technique is that the series resistance of the panel array containing 25 or 36 solar cells induces a power loss of 2% or 1.44%, respectively. Thus, the overall result of this finding is that ERE remains as the limiting factor for our multi-module PV arrays larger than 5 x 5.

Returning to Figure 7 above, in addition to 550K, there are many practical choices for the 1075-nm laser power and PV operating temperature. Figure 7 also shows that there are additional laser choices for Si-PV harvesting, such as the ~905 nm direct-diodes laser mentioned above. Our simulations (not shown here) reveal curves that are very close to those in Fig. 8. In other words, our diodes-laser simulations predict that when the 905 nm laser-beam illuminates the above-described solar panel, the resulting electric power outputs will be quite comparable to those for the Yb-fiber laser case.

**VIII. PV DIODE SUPPLEMENTED WITH TEG**

In recent years, several silicon-based TEG structures have been reported in the literature, where the active pillars are constructed from SiGe alloy or from GeSn alloy, a group-IV approach consistent with the silicon PV [16]. Also, thin-film TEG versions have been discussed as alternatives to the micro pillars [17]. In both cases, the area of the TEG can be large, and our suggestion is that the TEG area can match the area of the solar cell or panel. That being assumed, then we propose to use the GeSn TEG as a supplement to the Si PV. Figure 11 shows the hybrid or composite structure in which the TEG front face is in thermal contact with the PV rear face and the TEG rear face is in ambient air, yielding considerable temperature drop across the TEG as desired. In addition, the TEG literature reveals that the thermoelectric ZT figure of merit for the proposed Group IV TEG increases as the front face temperature of the TEG goes above 300K; for example, the $Ge_{0.86}Sn_{0.14}$ TEG with front face at 550K offers ZT = 0.92 [Fig. 6 of 18]

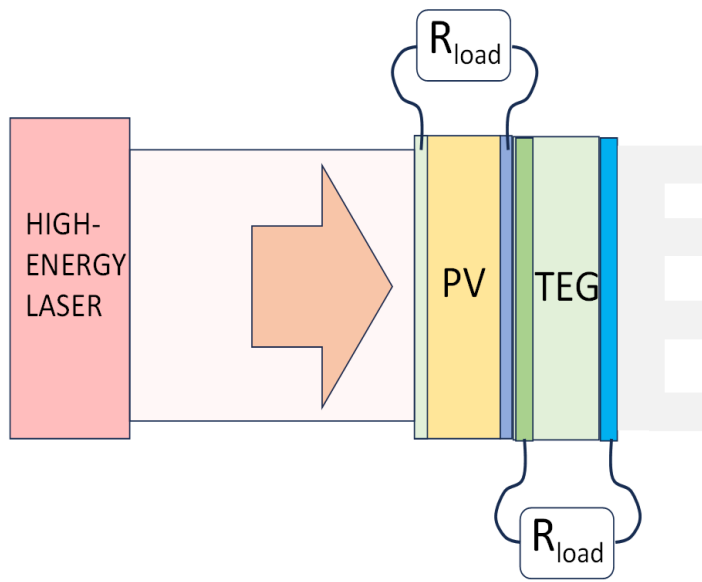

**Fig. 11.** Schematic drawing of the beam-actuated PV-and-TEG hybrid with combined electrical outputs.

This Fig.-11 approach provides two, combined sources of electric power, and is thereby superior to the Fig.-2 method. To give some context, Li *et al* [19] have proposed and analyzed a low-power laser-actuated GaAs-PV+TEG hybrid that is functionally similar to that proposed here. Leaving aside the laser, the literature reports that solar PV is enhanced by TEG [20]. Now we quantify the laser-power-to-electric-power

conversion efficiency of this hybrid. The hybrid system's efficiency is calculated as:

$$\eta_{PV-TEG} = \eta_{PV} + \eta_{TEG} \quad (7)$$

where, the maximum values of the $\eta_{TEG}$ is given by [20]:

$$\eta_{TEG}^{(max)} = \frac{T_h - T_c}{T_h}\left(\frac{\sqrt{1+ZT}-1}{\sqrt{1+ZT}+\frac{T_c}{T_h}}\right) \quad (8)$$

where $T_h$ and $T_c$ represent the hot and cold temperatures, respectively. As in Table II, the input power density provided by the laser is either 19013 W or 19388 W. Perfect heat transfer from PV to TEG is assumed in (7).

Table III summarizes the results for PV diode supplemented by TEG, with TEG based upon the GeSn-on-Si platform. If we look at the hybrid output electric power for the $Ge_{0.86}Sn_{0.14}$ TEG, we find 203 W output at the 19388 W input, a result that compares immediately with the Table-II result of 157 W for PV alone. This mean that hybrid gives up to 30% improvement in OE conversion. We say "up to 30%" because the Table-III hybrid with $Ge_{0.88}Sn_{0.12}$ TEG offers a 26% increase in total electrical output. Because Table III refers to an individual 15.6 cm x 15.6 cm PV-and-TEG cell, we shall now scale up the hybrid to the 5 x 5 array-of-cells "panel" discussed earlier for PV, and we shall then consider the resulting PV+TEG panel whose area is then 0.61 m², a factor-of-25 increase over Table III. If we then use the conservative estimate of a factor-of-20 increase in output electric power, the hybrid panel will provide 20 x 203 W, or about 4000 W as compared to the 3000 W found for the PV-only panel.

TABLE III. SUMMARY OF RESULTS PV+ TEG BASED UPON GROUP IV

| TEG System | Parameters | | |
|---|---|---|---|
| | ZT at 550 [K] | Efficiency ($\eta_{PV-TEG}$) $T_h$=550 [K]; $T_c$=278 [K] | Output electric power [W] For A=0.0243 m² |
| $Ge_{0.86}Sn_{0.14}$ | 0.92 | 0.42($d_{Si}$=180 μm); 0.43($d_{Si}$=500 μm) | 194($d_{Si}$=180 μm); 203($d_{Si}$=500 μm) |
| $Ge_{0.88}Sn_{0.12}$ | 0.80 | 0.41($d_{Si}$=180 μm); 0.42($d_{Si}$=500 μm) | 189($d_{Si}$=180 μm); 198($d_{Si}$=500 μm) |

Because the TEG panel is "expensive" (in some sense of the word), it is usually necessary to perform a "cost-benefits analysis" of the hybrid panel in order to determine whether the obtained increase in electric power is justified by the added costs of construction.

## IX. THE THERMORADIATIVE APPROACH

The TR diode definitely can be used during beaming to obtain useful amounts of electric power at its output terminals, but we have found that this approach is limited in the laser beaming context, and here are the details. Regarding direct laser-illumination of the TR, that is "forbidden" in the sense that band-to-band absorption reduces the TR conversion efficiency. Absorption would create an unwanted photocurrent flowing in opposition to the desired electric current. In other words, the photon-absorption current is a loss current reducing the output electric current. The TR approach is attained by adding layers in thermal contact with the front face of the TR cell. A reflective layer is deposited on the TR input face, together with a thick layer that absorbs the laser beam in order to attain temperature rise in that absorber, an elevated temperature that is immediately transferred to the TR body.

Assuming those layers, then the question we are raising is whether the figure of merit h for TR is comparable to that given by the PV approach. The first question to be answered in TR simulation is the temperature of the absorber layer as it pertains to the incoming infrared power density. To be definite, we shall assume that this absorber is monocrystalline silicon, and shall then turn to the Figure-6 results above, which indicates absorber temperature rising into the 380-to-420 K range for $D_b = 10000$ W/m². Thus, we can quantify TR body temperature with $D_b$. Next, we turn to the theoretical TR results of Strandberg [21] who plots the output electrical-power density at the TR maximum power point (MPP) as a function of the bandgap $E_g$ of the semiconductor used in the TR diode (his Fig. 9). Electrical output power density curves for TR body temperatures of 500K, 750 K, and 1000K are presented assuming a TR radiative-face temperature of 300 to 400K. Looking at that Figure, we find that $E_g$ must be in the range of 0.1 to 0.3 eV in order to get high W/m² electrical outputs. We also see that semiconductors with $E_g > 1$ eV give extremely low output, orders of magnitude lower than that of narrow-gap materials, a result that rules out the use of silicon.

The group-IV alloy GeSn is an excellent choice for realizing the TR diode, and in particular we recommend the specific crystal alloy $Ge_{0.8}Sn_{0.2}$ in order to provide at bandgap of $E_g = 0.2$ eV at elevated temperature such as 550K. In addition, it is essential to point out that this TR semiconductor will become segregated or unstable at TR body temperatures above 550K. This then places an upper limit on the TR operation temperature [21]. In particular, we see in [21] an ideal TR output power density of 200 W/m² at 550K. If we compare that density with our Table II PV density result, we see that TR is 30x smaller, which is why PV is primary here.

## X. SYSTEM COSTS and BENEFITS

Regarding the directed-energy system costs, these include maintenance costs, the capital cost of the laser, the cost of fueling or "powering" the laser, the cost of moving the laser (by mounting it on a truck, for example), the costs of the laser-aiming system (including real-time tracking when the energy beaming is required for moving systems) the costs of the PV cell, and the cost of its associated electrical circuitry. We are not saying that these costs are low. In fact, the overall cost could be high. We are saying that paying the total cost will be justified in most cases by the unique and valuable capabilities of the new electrification system. The benefits of the system will make the financial investment worthwhile.

## XI. III-V SEMICONDUCTOR PHOTOVOLTAICS

We wish to present a wider context for our silicon-solar approach by detailing the excellent progress that has been made on III-V semiconductor photovoltaic cells during the last several years. A series of experiments on InGaAs, InGaAsP, GaAs, and GaSb PV devices, both single-junction and multi-junction devices, has proven the value of these devices for harvesting electric power from the beam of a Nd:YAG laser emitting at the 1064-nm wavelength [22-27]. The laser beam power incident upon the converter was in the range 0.5 to 4 W CW. Optical input power of up to 50 W CW from a 980-nm diode laser was also investigated using III-Vs. By selecting their 300K bandgaps to be below 1.12 eV (the Si $E_g$), it is clear that the InGaAs and InGaAsP PVs will quite successfully convert the laser beam from the Yb-doped-fiber lasers and direct-diodes lasers that are targeted in this paper. Therefore, the III-Vs definitely provide an alternative to silicon in the beaming system.

Regarding the power conversion efficiency (PCE) of the InGaAs PV at 1064 nm, two authors have projected that PCE decreases significantly as the PV cell temperature is increased to 373K. The efficiency's rate-of-decrease is shown in Fig. 5 of [22] and in Fig. 7(d) of [24]. Extrapolating the InGaAs cell temperature to 550K, we find the PCE falling to 16% for the multi-junction device [22].

To assist the "choice of semiconductor" for the PV device, we shall now sketch a silicon-to-III-V comparison by considering technical factors (such as thermal management during UHP illumination) and the manufacturing cost factors.

On a theoretical basis, we can compare the "thermal performance" of the InGaAs PV panel to that of the Si PV panel, assuming for both panels the same illumination area, the same heat sink structure, the same 1075-nm laser operating wavelength, the same incident laser power such as 10 kW CW, and the same bandgap obtained by adjusting III-V alloy composition (giving similar-to-Si absorption spectra). We note that all of the InGaAs PV cells reported thus far have an overall layered thickness of 10 μm or less, whereas the Si PN cell thickness is in the 180-to-500 μm range. Because the InGaAs panel is comparatively "thin", our thermal analysis projects that the III-V panel will settle at a higher steady-state temperature than the Si does. If that is correct, then if we impose the same maximum operating temperature $T_m$ upon both panels, the III-V will reach that $T_m$ at a laser power input that is less than that for Si; hence the III-V gives less electrical output than the Si offers.

Turning to the manufacturing and production of PV cells and PV panels, we note three factors: (1) the construction of the III-V PVs is generally more complex than that of Si PVs, (2) the PV cell size in the III-V case will be generally smaller than the Si cell size due to the smaller diameter of available InP and GaAs "substrate" wafers as compared to the Si wafer diameter, a fact that may make the III-V PV panel assembly more difficult, and (3) the various cost factors appear to be larger in the III-V case.

## XI. SUMMARY

In the introduction to this paper, it is stated that the optimized electrical output occurs at a PV temperature well above room temperature, and the statement seems to contradict the well-known decrease in PV efficiency with increased temperature. However, there is no contradiction because the electric output is given by the *product* of the efficiency and the laser power inputted to the PV. In the present system, as PV temperature rises up from 293K towards 600K, the rate of laser power increase is slightly larger than the rate of efficiency decrease, and for that reason, the electric output power is maximum at the maximum allowed temperature.

We can outline the contribution of this paper by noting that the paper proposes, and gives quantitative theoretical analysis of, a novel laser-driven optical-to-electrical "power by light" system in which the innovative components of the system work together in a synergistic way to produce considerable electric power at a remote location. The multi-kilowatt electrical outputs that are predicted are based upon thermal, optical, and electrical modeling-and-simulation. Eleven scenarios for practical application of the directed-energy beaming over kilometers distance with low loss through the atmosphere are presented. The novel aspects of the system are: (1) utilization of ultra-high-power CW SWIR laser beams giving 20 kW of power, (2) silicon photovoltaic OE conversion cells that are commercial solar cells "repurposed" for UHP monochromatic light, (3) large-area panels comprised of horizontally interconnected PV cells that "harvest" effectively after reaching a stable panel temperature during 20 kW/m2 illumination by a collimated beam, (4) passive heat sinking of the panel instead of an electrically powered heat sink that deploys blowers and pumped liquids, (5) operation of PV cells and panels at elevated temperatures around 550K, as discussed above, (6) series electrical connection of all cells in the panel to ensure full undiminished electrical output, (7) comparative analysis of thermo-radiative cells for beam conversion, (8) proposed panels comprised of hybrid PV + TEG modules for enhanced electrical output, and (9) a comparison of the Si PV converter with the InGaAs and InGaAsP photovoltaic approaches, suggesting that thermal management of the III-V PVs is problematic.

We can also sketch the strengths and weakness of the present system as compared to existing systems in the literature. The present strengths are: (1) the proposed system builds upon widely proven and widely adopted solar panel technology, (2) the system builds upon widely proven UHP laser technology, (3) the scope of potential practical applications is wide. Regarding weaknesses, the list includes: (1) the system cost is high, (2) the overall energy efficiency might be low when the energy needed to supply the laser is taken into account, (3) there are eye safety issues at the sending and receiving stations, (4) the beaming could be interrupted by smog, fog, smoke and rain, (5) a pair of lenses is required at the laser station to form the collimated beam with desired diameter, (6) the high-temperature array might present some danger of burns or fire. It is important to note that all six of these factors are not specific to our system. They are six generic aspects that apply to any UHP beaming system, which means that these weaknesses are anticipated regardless of the hardware that is used for the PV panel, the laser supply, the laser-directing apparatus, etc.

Compared to existing power-by-light systems, the present system handles beam powers that are three orders-of-magnitude higher than those of existing systems. Also, in most existing systems, the beam is focused to a spot on the PV converter, with the spot having a diameter of a few centimeters.

## XII. CONCLUSIONS

We have presented potential applications of UHP CW laser beaming to distant places where the harvesting of that beam by a semiconductor photodetector provides considerable electric power, power that is beneficial to various users and to society generally. In principle, a thermo-radiative (TR) diode could be employed for harvesting, but our studies indicate a TR electrical output that is much lower than that provided by a photovoltaic (PV) cell, thereby making PV the primary means.

For the UHP Yb-doped 1075-nm fiber laser, it is a fortuitous coincidence that the silicon solar cell is an ideal optical-to-electrical converter, and that the cell can be a well-known commercial cell. Because the laser beam can be delivered through the atmosphere with very low loss, and because the collimated beam can have a diameter of around one meter, a silicon solar-cell "panel" can be deployed for efficient harvesting. We have performed here a series of thermal and optical simulations that quantify the performances that can be expected; for example, performance using a monocrystalline silicon layer of thickness in the 180 to 500 μm range. Our results indicate that 500 μm is better than 180 μm, but not much better. Our results also predict about 15% OE conversion in the laser power range of 10 kW to 20 kW, with panel temperature in the 436K to 560K range--in particular, an electrical output of 3000 W from a 0.6 m$^2$ panel illuminated by 20 kW 1075-nm beam, where the panel operates at a temperature of 550 K.

To obtain additional electrical power output, the PV cell can be supplemented by a TEG cell whose area matches the PV area, and this hybrid uses thermal contact between the PV rear face and the TEG front face.

**Acknowledgment**

R.S. acknowledges the support of the U.S. Air Force Office of Scientific Research on Grant No. FA9550-21-1-0347. O.M. acknowledges support from NSERC Canada (Discovery, SPG, and CRD Grants), Canada Research Chairs, Canada Foundation for Innovation, Mitacs, PRIMA Québec, Defence Canada (Innovation for Defence Excellence and Security, IDEaS), the European Union's Horizon Europe research and innovation program under grant agreement No 101070700 (MIRAQLS), the U.S. Army Research Office on Grant No. W911NF-22-1-0277, and the U.S. Air Force Office of Scientific and Research on Grant No. FA9550-23-1-0763.

## AUTHOR DECLARATION

**Conflict of Interest**

The authors have no conflicts to disclose.

**Author Contributions**

**Richard Soref,** Conceptualization (lead); Project administration (equal), Supervision (equal), Writing- original draft (lead), Writing- review & editing (equal), **Francesco De Leonardis,** Conceptualization (supporting), Formal analysis (lead), Software (lead), Writing- original draft (supporting), Writing- review & editing (equal), **Gerard Daligou,** Formal analysis (supporting), Writing- review & editing (equal), **Oussama Moutanabbir,** Conceptualization (supporting), Project administration (equal), Supervision (equal), Writing- review & editing (equal).